# Highly oriented SrTiO$_3$ thin film on graphene substrate

*Sang A Lee[†,‡,⊥], Jae-Yeol Hwang[§,⊥], Eun Sung Kim[§], Sung Wng Kim[*,∥], and Woo Seok Choi[*,†]*

[†]Department of Physics, Sungkyunkwan University, Suwon, 16419, Korea

[‡]Institute of Basic Science, Sungkyunkwan University, Suwon, 16419, Korea

[§]Center for Integrated Nanostructure Physics, Institute for Basic Science (IBS), Suwon 16419, Korea

[∥]Department of Energy Sciences, Sungkyunkwan University, Suwon 16419, Korea

KEYWORDS: SrTiO$_3$, graphene, thin film, pulsed laser deposition, interface, epitaxy, diffusion barrier, Si

ABSTRACT: Growth of perovskite oxide thin films on Si in crystalline form has long been a critical obstacle for the integration of multifunctional oxides into Si-based technologies. In this study, we propose pulsed laser deposition of a crystalline SrTiO$_3$ thin film on a Si using graphene substrate. The SrTiO$_3$ thin film on graphene has a highly (00$l$)-oriented crystalline structure which results from the partial epitaxy. Moreover, graphene promotes a sharp interface by highly suppressing the chemical intermixing. The important role of graphene as a 2D substrate and diffusion barrier allows expansion of device applications based on functional complex oxides.

Functional perovskite oxides exhibit a variety of emergent physical phenomena such as superconductivity, 2D electron liquid behavior, multiferroicity, catalytic activity, and thermoelectricity, and have high potential for application to opto-electronic and energy devices.[1-5] One of the main obstacles for the practical utilization of such functionalities of perovskite oxides is their compatibility with the existing Si-based technology. In other words, high-quality single-crystalline epitaxial perovskite oxide thin films and heterostructures are grown almost exclusively on single-crystalline perovskite substrates, *e.g.*, $SrTiO_3$, owing to the structural and chemical affinities of the perovskite building blocks.

One of the overarching goals for enabling the practical application of functional oxides is, therefore, to grow single-crystalline $SrTiO_3$ thin films on Si.[6-12] The $SrTiO_3$ layer serves as a universal template for the growth of most perovskites, and thus, various functional oxide systems might become readily compatible with Si-based electronics. However, the growth of single-crystalline $SrTiO_3$ thin films on Si is not trivial, since most conventional growth procedures on a Si substrate result in a polycrystalline or an amorphous phase of $SrTiO_3$. The difficulty in the thin film growth arises mainly from the thermal mismatch between Si and oxides and the chemical stability of Si in their vicinity.[11] Even though the lattice mismatch (a compressive strain of ~1.7%) between $SrTiO_3$ and Si is rather small,[10] there is a significant discrepancy in their thermal expansion coefficients. More importantly, Si becomes chemically unstable over 400°C, resulting in the formation of a detrimental intermediate layer such as amorphous $SiO_2$ ($a$-$SiO_2$) or silicides ($SrSi_2$ and/or $TiSi_2$) at the oxide/Si interface.[7, 11, 13-15] In an attempt to minimize the influence of the interface and enhance the crystallinity of the $SrTiO_3$ thin film, previous studies employed complicated approaches such as stepped growth, ultraviolet ozone treatment, and the use of high-flux sources.[7, 11-12, 16] Most of these examples were based on molecular beam epitaxy (MBE) which was quite successful in obtaining single-crystalline $SrTiO_3$. Nevertheless, the use of more accessible growth techniques such as sputtering and pulsed laser deposition (PLD) can also be envisaged. While PLD has some drawbacks such as limitations in the sample size, the growth of single-crystalline $SrTiO_3$ thin films on Si could economically expand the study of complex oxide materials systems on Si.

Recently, graphene has been recognized as a 2D layer that facilitates the growth of crystalline layers.[17-19] Graphene is a 2D monolayer of carbon atoms in a honeycomb lattice structure, and it promotes van der Waals epitaxy to materials such as GaAs and other 2D layered materials.[17] Due to this weak van der Waals interaction instead of typical strong covalent or ionic bonding, discrepancies in the lattice constant and crystalline symmetry between the thin film and graphene substrate are much less prominent than those in a film deposited on conventional substrates such as $SiO_2$/Si.[20] Moreover, graphene is chemically and thermally stable, because of which it serves as an ideal substrate for epitaxial and/or single-crystalline thin film growth. Finally, given the advantage of graphene to be transferred to any arbitrary substances, including amorphous $SiO_2$ and glass, graphene substrate can facilitate the realization of ubiquitous applications of high-quality single-crystal materials. Such advantageous characteristics of graphene as a 2D substrate allow us to explore the growth of single-crystalline perovskite oxides on graphene.

In this paper, we report the growth of a highly oriented SrTiO$_3$ thin film on Si by using graphene as the 2D substrate. Despite the large lattice and symmetry mismatches, the SrTiO$_3$ thin film shows a highly (00$l$)-oriented crystal structure on the graphene transferred onto SiO$_2$/Si substrate. In addition, the graphene layer promotes a sharp interface without the presence of a detrimental intermediate layer, which suggests that the graphene also serves as a suitable diffusion barrier.

In order to examine the suitability of graphene as a 2D substrate for crystalline complex oxides, we used PLD to grow SrTiO$_3$ thin films on graphene-covered $a$-SiO$_2$/Si substrates. The chemical-vapor-deposition-grown graphene was transferred onto $a$-SiO$_2$/Si (001) substrate. PLD was subsequently used to grow SrTiO$_3$ thin films on graphene substrate at 700°C. Low oxygen partial pressure (~10$^{-6}$ Torr) was employed in order to prevent oxidization of the graphene layer. For comparison, SrTiO$_3$ thin films were also deposited on a bare $a$-SiO$_2$/Si substrate at the same time.

Structural characterization of the SrTiO$_3$ thin film on graphene substrate confirmed a crystalline structure with predominant (00$l$) orientation, as shown schematically in Figure 1a. Figure 1b shows x-ray diffraction (XRD) $\theta$-$2\theta$ scans and x-ray reflectivity (XRR) results of the SrTiO$_3$ thin film grown on the $a$-SiO$_2$/Si substrate with (red) and without (blue) graphene. The XRD result of the SrTiO$_3$ thin film on graphene shows clear (00$l$) SrTiO$_3$ peaks, indicating a highly (00$l$)-oriented thin film. While weak signatures of (110), (111), and (211) SrTiO$_3$ peaks (at $2\theta$ = 32.50°, 40.03°, and 57.84°, respectively) are also present indicating that the film is not fully single-crystalline, there is a stark distinction between the SrTiO$_3$ thin film on graphene and $a$-SiO$_2$/Si. Indeed, the thin film on $a$-SiO$_2$/Si without graphene does not suggest any crystalline structures, except for miniscule hints of (110), (111), (002), and (211) SrTiO$_3$ peaks. Instead, the wide bump at low $2\theta$ angles indicates that, as predicted, the film is largely amorphous on Si. The XRD $\phi$-scan shown in the inset of Figure 1b further supports the local ordering of SrTiO$_3$ thin film with four-fold symmetry on graphene substrate. The XRR results also reveal a strong influence of the graphene layer on the growth of the SrTiO$_3$ thin film. Clear interference fringes are observed for SrTiO$_3$ on graphene at grazing incident angles indicating a well-defined SrTiO$_3$-graphene interface and the SrTiO$_3$ surface compared to the sample without the graphene layer, most probably due to the substantially reduced intermixing at the interface.

Transmission electron microscopy (TEM) further confirms the highly oriented crystalline nature of the SrTiO$_3$ thin film on graphene substrate. The cross-sectional TEM image of the SrTiO$_3$ thin film with the graphene layer (Figure 2a) shows (00$l$)-oriented columnar texture with the local epitaxial growth of SrTiO$_3$, which is in good agreement with the XRD results. The electron diffraction pattern for the well-ordered region (orange box, Figure 2c) exhibits clear spots in the reciprocal lattice, indicating a single-crystalline feature with [00$l$] orientation. For another region, twin structures are observed (blue box, Figure 2d). Nevertheless, all the grains are oriented along the [00$l$] direction. The high-resolution image in Figure 2e with the intensity line profile of the A- and B-site cations confirms the formation of the expected perovskite structure. In contrast, in the case of the SrTiO$_3$ thin film grown directly on $a$-SiO$_2$/Si, atomic orderings are observed only partially, indicating an amorphous phase. We further analyzed magnified TEM images near the interface, as shown in Figures 2f and 2g. The SrTiO$_3$ thin film on graphene substrate shows clear interface and well-arranged atomic structure. The termination layer can be roughly estimated as TiO$_2$, which coincides with our expectation as discussed below. On the other hand, the SrTiO$_3$ thin film without graphene shows largely ambiguous interface.

The coherent growth of the SrTiO$_3$ thin film on graphene might seem rather surprising at first sight, especially considering the mismatch in crystalline symmetry between the film and substrate. In particular, the (111) surface of SrTiO$_3$ seems to be epitaxially more favorable for being attached on top of the honeycomb structure of graphene, due to its hexagonal symmetry. However, this is not the case apparently, because of the following reasons. First, the lattice mismatch for epitaxial growth is actually greater for (111)-oriented SrTiO$_3$ than for (001)-oriented SrTiO$_3$ on graphene. A lattice mismatch calculation reveals that the epitaxial strain induced by graphene on the (111) SrTiO$_3$ thin film is ~11.23%, whereas it reduces to 8.72% (armchair direction) and −5.40% (zigzag direction) for the (001) SrTiO$_3$ thin film if we consider a simple square box on top of the hexagonal network (see Figure S1 in Supporting Information). Second, the surface energy ($E^S$) is lowest for the (001) surface of SrTiO$_3$. According to a recent density functional theory calculation, $E^S$ (100) ($\approx$ 6.8 eV/nm$^2$) is the lowest among the surface energies of low Miller-index surfaces, by a factor of at least three.[21] When ablated species arrive at heated substrates, atoms tend to form a new surface with the lowest surface energy. Third, an atomic termination layer for the (001) SrTiO$_3$ surface (TiO$_2$ layer) has higher adsorption and migration energies on graphene compared to that for the (111) surface (Ti-only layer). In particular, oxygen (4.79 eV) has higher adsorption energy than does Ti (3.27 eV), which suggests that it is more favorable to form a TiO$_2$ layer on top of graphene instead of a Ti-only layer. On the other hand, Sr has the lowest adsorption energy (0.33 eV) among the elements, suggesting that the Ti-containing layer should be attached on graphene.[22] This is in agreement with our TEM observation.

The discussion on the bonding nature and epitaxial strain provides further insight into the growth of the highly oriented SrTiO$_3$ thin film on graphene substrate. We first demonstrate that the bonding between graphene and SrTiO$_3$ is not purely a van der Waals interaction. The adsorption energy discussed above can promote a stronger bonding nature between the carbon and Ti/oxygen ions, which can induce an epitaxial strain. While the detailed nature of the chemical bonding is beyond the scope of the present study, the evidence of the epitaxial strain can be found from the structural characterizations. The out-of-plane lattice constant of the SrTiO$_3$ layer calculated from the XRD (00$l$) peaks is ~3.896 Å (Figure 1b). This value is in close agreement with the TEM results shown in Figure 2e. On the other hand, the in-plane lattice constant of the SrTiO$_3$ thin film is ~3.923 Å, from an off-axis XRD measurement. The bulk lattice constant of single-crystal SrTiO$_3$ is 3.905 Å, indicating a tensile-strained SrTiO$_3$ thin film on graphene.[23] It should further be noted that the lattice constant calculated from XRD peaks other than the main (00$l$) peaks is almost consistent with the bulk value. This indicates that the highly oriented main part of the SrTiO$_3$ thin film is epitaxially strained, whereas other small disoriented regions might be partially relaxed. The moderate strength of the bonding between 2D graphene and the SrTiO$_3$ thin film seems to promote the highly oriented crystalline growth by overcoming the symmetry and lattice mismatches.[17] Another possible mechanism of the partial epitaxial growth notwithstanding with the previous discussion is thermal mismatch between SrTiO$_3$ and graphene.[24] The thermal expansion coefficients of SrTiO$_3$, graphene, and Si are $9.0 \times 10^{-6}$ K$^{-1}$, $-8.0 \times 10^{-6}$ K$^{-1}$ and $2.5 \times 10^{-6}$ K$^{-1}$, respectively, exhibiting a large difference.

In addition to facilitating the partially epitaxial growth of SrTiO$_3$, the graphene layer efficiently limits the intermixing of elements during the highly energetic PLD. While such an effect has been demonstrated within perovskite oxide heterostructures,[25] the effect shown in the present study is much more pronounced owing to the structural robustness of graphene. As shown in Figure 2a, the SrTiO$_3$ thin film grown on graphene substrate has a sharp interface, which is also confirmed from the XRR result (Figure 1b). In contrast, for the thin film without graphene, it is rather difficult to identify the interface between the SrTiO$_3$ thin film and the *a*-SiO$_2$ /Si substrate, as shown in Figure 2b. This indicates severe elemental intermixing at the interface during the high temperature deposition. Furthermore, the obscure interface adversely influences the lattice arrangement of the entire thin film and results in the formation of an amorphous phase of the SrTiO$_3$ layers. The observation of a well-defined interface in the SrTiO$_3$ thin film on graphene substrate should be also beneficial for controlling the physical properties, since the possible SiO$_x$ layer is subject to direct tunneling leakage and poor reliability.[26] This result indicates that the graphene layer plays the role of a robust diffusion barrier without requiring complicated approaches for the stabilization of the heterointerface. The use of graphene substrate within PLD is comparable to the Sr layer engineering in MBE, which also reduces the intermixing.[27]

Figure 3 shows the Raman spectrum of the SrTiO$_3$ grown on graphene. The spectra of the SrTiO$_3$ thin film without graphene and that of pristine graphene transferred onto a Si substrate are also shown for comparison. First, we observe the Raman peaks originating from the graphene layer for the SrTiO$_3$ on graphene, although both G (~1600 cm$^{-1}$) and 2D (~2680 cm$^{-1}$) peaks are rather broad. In addition, we notice D peak (~1360 cm$^{-1}$), which can be possibly attributed to the disorder induced by the growth of the SrTiO$_3$ thin film at high temperature. The 2D peak for the SrTiO$_3$ on graphene shows a slight blue-shift compared to the bare graphene. This shift might be due to a combined effect of strain and oxidation in graphene layer.[28] Three peaks originating from the SrTiO$_3$ thin film (indicated by yellow arrows) are also visible: TO$_1$ (~100 cm$^{-1}$), LO$_4$ (~800 cm$^{-1}$), and an unidentified peak at ~1360 cm$^{-1}$.[29] The peak at ~1360 cm$^{-1}$, which overlaps with the D peak of graphene, has been previously reported for the SrTiO$_3$ thin film, although it has not been attributed to any specific phonon mode.[30] The presence of the Raman peaks and the enhancement of the LO$_4$ peak originating from the $SrTiO_3$ thin film on graphene substrate in comparison to that from the $SrTiO_3$ thin film without graphene suggest the successful formation of the $SrTiO_3$ layer on top of graphene substrate.

In summary, pulsed laser deposition of a highly-orientated $SrTiO_3$ thin film with a sharp interface on Si was demonstrated via the introduction of graphene as a 2D substrate and a diffusion barrier. The moderate strength of the bonding between the carbon and $SrTiO_3$ layers facilitated the growth of the predominantly (00*l*)-oriented $SrTiO_3$ thin film on graphene substrate. The bonding between graphene and the $SrTiO_3$ thin film induced the partial epitaxial growth generating an in-plane tensile strain, which was evidenced from various structural analyses. Given the extraordinary structural features of our pulsed laser deposited $SrTiO_3$ thin film on graphene substrate, we expect better integration of functional transition metal oxides into Si-based technology.

FIGURES

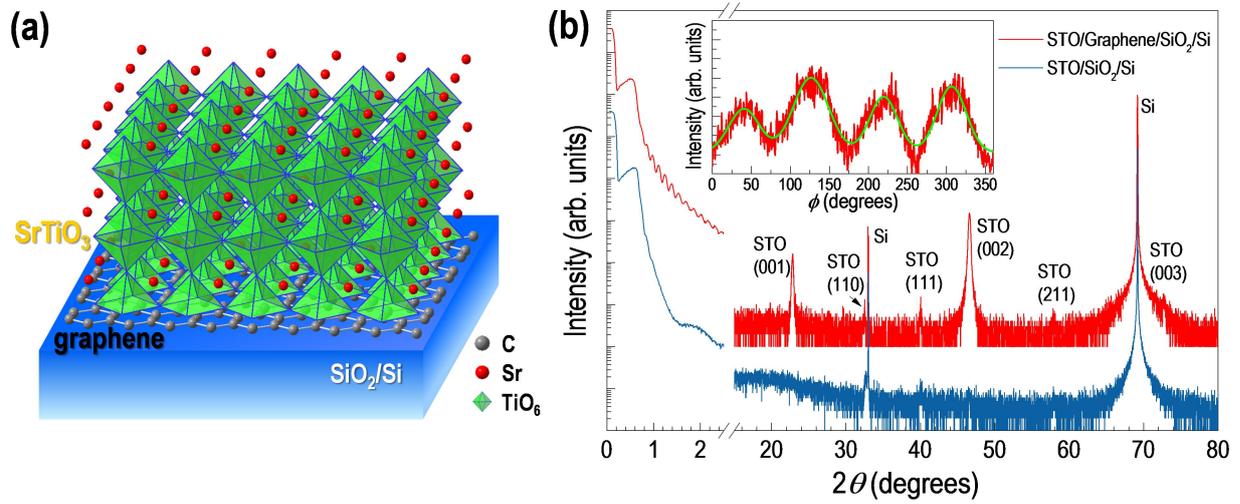

**Figure 1.** Growth of highly oriented SrTiO$_3$ thin film on 2D graphene substrate. (a) A schematic representation of the (00*l*)-oriented perovskite SrTiO$_3$ thin film on graphene substrate. (b) X-ray diffraction results of SrTiO$_3$ thin films grown on *a*-SiO$_2$/Si substrates with and without graphene show a substantial difference in crystal structures. For the film on graphene (red curve), strong (00*l*) Bragg reflection peaks of SrTiO$_3$ are clearly observed, indicating the highly oriented SrTiO$_3$ thin film. On the other hand, the film without graphene does not show any significant diffraction peak indicating a largely amorphous structure. The inset shows XRD $\phi$-scan result of SrTiO$_3$ thin film on graphene substrate, for the 103 reflection.

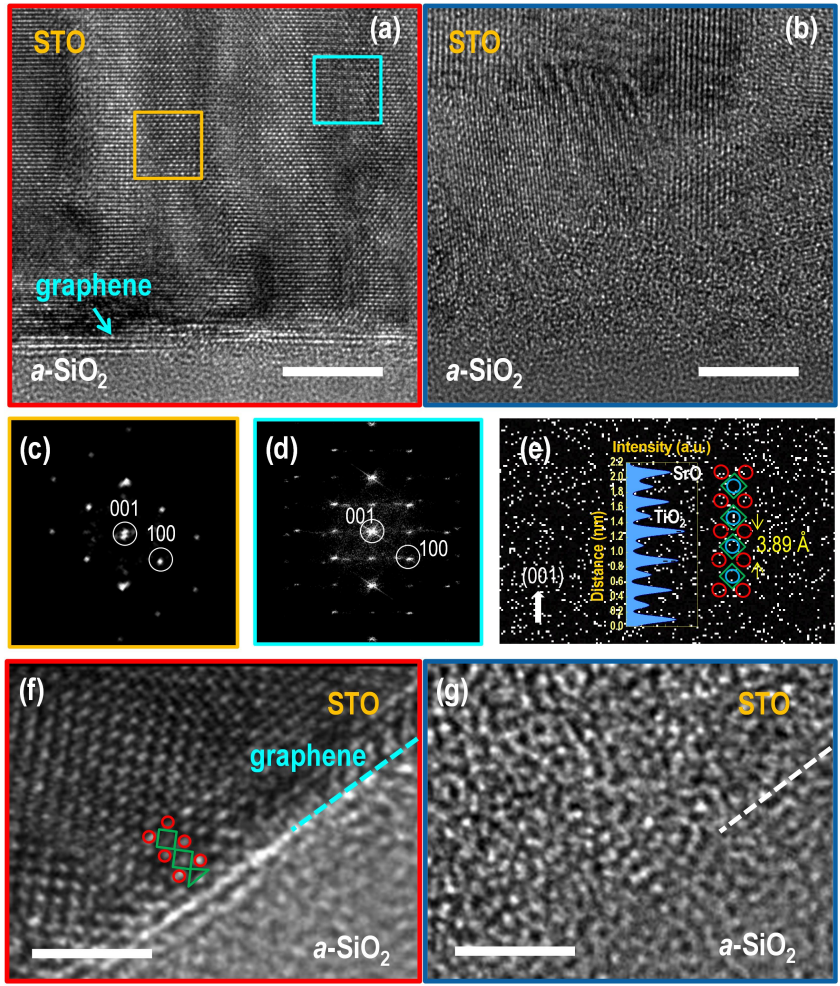

**Figure 2.** Transmission electron microscopy images of SrTiO$_3$ thin films with and without graphene. The SrTiO$_3$ thin films (a) with and (b) without the graphene show crystalline and amorphous structures, respectively, which is consistent with the XRD results. The film with graphene shows a well-defined interface owing to the structural integrity of graphene. The electron diffraction patterns in the (c) orange and (d) blue boxes show single-crystalline and twin structures, respectively, within the images of the SrTiO$_3$ thin film on graphene substrate. (e) Magnified image of the single-crystalline region. The line profile shows that the *c*-axis lattice parameter is 3.89 Å, which is in good agreement with the XRD results. Magnified images of SrTiO$_3$ thin films (f) with and (g) without graphene around the interface. The scale bars represent 5 nm (a and b) and 2 nm (f and g), respectively.

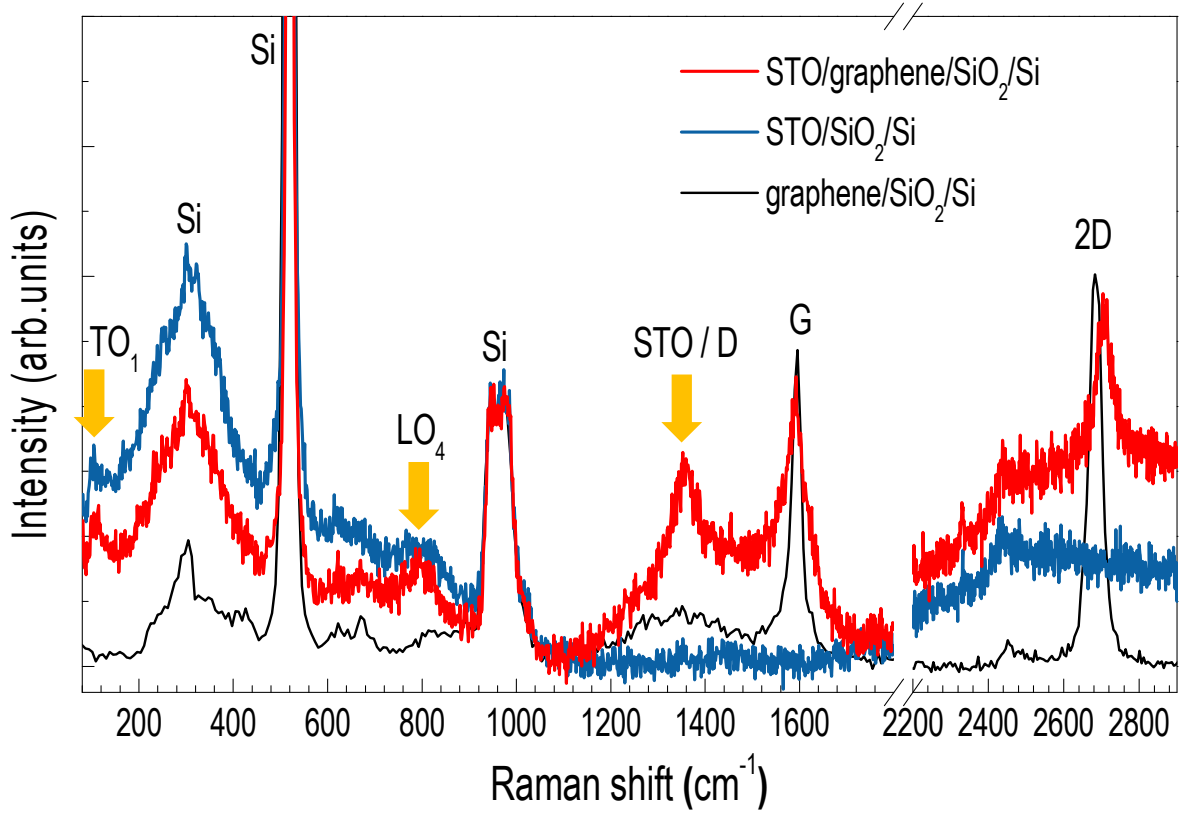

**Figure 3.** Phonon spectra analyses of SrTiO$_3$ thin films on graphene. The Raman spectra of the SrTiO$_3$ thin films on the *a*-SiO$_2$/Si substrate with (red) and without (blue) graphene show strong SrTiO$_3$ phonon peaks. The arrows indicate the peaks originating from the SrTiO$_3$ phonon. All the spectra are normalized to the Si peaks at ~960 cm$^{-1}$.

## ASSOCIATED CONTENT

This material is available free of charge via the Internet at http://pubs.acs.org.

Experimental procedures and a lattice mismatch calculation of (100)- and (111)-oriented SrTiO$_3$ on graphene substrate and temperature-dependent sheet resistance of SrTiO$_3$ thin films (PDF)

## AUTHOR INFORMATION

### Corresponding Author

* E-mail: kimsungwng@skku.edu (S.W.K.)

* E-mail: choiws@skku.edu (W.S.C.)

### Author Contributions

$^\perp$These authors contributed equally to this work.

### Notes

The authors declare no competing financial interest.

## ACKNOWLEDGMENT

This work was supported by Basic Science Research Programs through the National Research Foundation of Korea (NRF) (NRF-2014R1A2A2A01006478 (W.S.C.) and NRF-2013R1A1A2057523 (S.A L.)). This work was also supported by IBS-R011-D1 (J.-Y.H., E.S.K., S.W.K., and W.S.C.).

REFERENCES

(1)     Hwang, H. Y.; Iwasa, Y.; Kawasaki, M.; Keimer, B.; Nagaosa, N.; Tokura, Y., Emergent Phenomena at Oxide Interfaces. *Nat. Mater.* **2012,** *11*, 103-113.

(2)     Zubko, P.; Gariglio, S.; Gabay, M.; Ghosez, P.; Triscone, J.-M., Interface Physics in Complex Oxide Heterostructures. *Annu. Rev. Condens. Matter Phys.* **2011,** *2*, 141-165.

(3)     Wang, J.; Neaton, J. B.; Zheng, H.; Nagarajan, V.; Ogale, S. B.; Liu, B.; Viehland, D.; Vaithyanathan, V.; Schlom, D. G.; Waghmare, U. V.; Spaldin, N. A.; Rabe, K. M.; Wuttig, M.; Ramesh, R., Epitaxial $BiFeO_3$ Multiferroic Thin Film Heterostructures. *Science* **2003,** *299*, 1719-1722.

(4)     Suntivich, J.; May, K. J.; Gasteiger, H. A.; Goodenough, J. B.; Shao-Horn, Y., A Perovskite Oxide Optimized for Oxygen Evolution Catalysis from Molecular Orbital Principles. *Science* **2011,** *334*, 1383-1385.

(5)     Ohta, H.; Kim, S.; Mune, Y.; Mizoguchi, T.; Nomura, K.; Ohta, S.; Nomura, T.; Nakanishi, Y.; Ikuhara, Y.; Hirano, M.; Hosono, H.; Koumoto, K., Giant Thermoelectric Seebeck Coefficient of a Two-Dimensional Electron Gas in $SrTiO_3$. *Nat. Mater.* **2007,** *6*, 129-134.

(6)     McKee, R. A.; Walker, F. J.; Chisholm, M. F., Crystalline Oxides on Silicon: The First Five Monolayers. *Phys. Rev. Lett.* **1998,** *81*, 3014-3017.

(7)     Choi, M.; Posadas, A.; Dargis, R.; Shih, C.-K.; Demkov, A. A.; Triyoso, D. H.; David Theodore, N.; Dubourdieu, C.; Bruley, J.; Jordan-Sweet, J., Strain Relaxation in Single Crystal $SrTiO_3$ Grown on Si (001) by Molecular Beam Epitaxy. *J. Appl. Phys.* **2012,** *111*, 064112.

(8)     Niu, G.; Penuelas, J.; Largeau, L.; Vilquin, B.; Maurice, J. L.; Botella, C.; Hollinger, G.; Saint-Girons, G., Evidence for the Formation of Two Phases During the Growth of $SrTiO_3$ on Silicon. *Phys. Rev. B* **2011,** *83*, 054105.

(9)     Kolpak, A. M.; Walker, F. J.; Reiner, J. W.; Segal, Y.; Su, D.; Sawicki, M. S.; Broadbridge, C. C.; Zhang, Z.; Zhu, Y.; Ahn, C. H.; Ismail-Beigi, S., Interface-Induced Polarization and Inhibition of Ferroelectricity in Epitaxial SrTiO$_3$/Si. *Phys. Rev. Lett.* **2010,** *105*, 217601.

(10)    Warusawithana, M. P.; Cen, C.; Sleasman, C. R.; Woicik, J. C.; Li, Y.; Kourkoutis, L. F.; Klug, J. A.; Li, H.; Ryan, P.; Wang, L.-P.; Bedzyk, M.; Muller, D. A.; Chen, L.-Q.; Levy, J.; Schlom, D. G., A Ferroelectric Oxide Made Directly on Silicon. *Science* **2009,** *324*, 367-370.

(11)    Reiner, J. W.; Kolpak, A. M.; Segal, Y.; Garrity, K. F.; Ismail-Beigi, S.; Ahn, C. H.; Walker, F. J., Crystalline Oxides on Silicon. *Adv. Mater.* **2010,** *22*, 2919-2938.

(12)    Niu, G.; Saint-Girons, G.; Vilquin, B.; Delhaye, G.; Maurice, J.-L.; Botella, C.; Robach, Y.; Hollinger, G., Molecular Beam Epitaxy of SrTiO$_3$ on Si (001): Early Stages of the Growth and Strain Relaxation. *Appl. Phys. Lett.* **2009,** *95*, 062902.

(13)    Forst, C. J.; Ashman, C. R.; Schwarz, K.; Blochl, P. E., The Interface between Silicon and a High-k Oxide. *Nature* **2004,** *427*, 53-56.

(14)    Hu, X.; Li, H.; Liang, Y.; Wei, Y.; Yu, Z.; Marshall, D.; Edwards, J.; Droopad, R.; Zhang, X.; Demkov, A. A.; Moore, K.; Kulik, J., The Interface of Epitaxial SrTiO$_3$ on Silicon: In Situ and Ex Situ Studies. *Appl. Phys. Lett.* **2003,** *82*, 203-205.

(15)    Goncharova, L. V.; Starodub, D. G.; Garfunkel, E.; Gustafsson, T.; Vaithyanathan, V.; Lettieri, J.; Schlom, D. G., Interface Structure and Thermal Stability of Epitaxial SrTiO$_3$ Thin Films on Si (001). *J. Appl. Phys.* **2006,** *100*, 014912.

(16)    Li, H.; Hu, X.; Wei, Y.; Yu, Z.; Zhang, X.; Droopad, R.; Demkov, A. A.; Edwards, J.; Moore, K.; Ooms, W.; Kulik, J.; Fejes, P., Two-Dimensional Growth of High-Quality Strontium Titanate Thin Films on Si. *J. Appl. Phys.* **2003,** *93*, 4521-4525.

(17) Alaskar, Y.; Arafin, S.; Wickramaratne, D.; Zurbuchen, M. A.; He, L.; McKay, J.; Lin, Q.; Goorsky, M. S.; Lake, R. K.; Wang, K. L., Towards van der Waals Epitaxial Growth of GaAs on Si using a Graphene Buffer Layer. *Adv. Funct. Mater.* **2014,** *24*, 6629-6638.

(18) Shi, Y.; Zhou, W.; Lu, A.-Y.; Fang, W.; Lee, Y.-H.; Hsu, A. L.; Kim, S. M.; Kim, K. K.; Yang, H. Y.; Li, L.-J.; Idrobo, J.-C.; Kong, J., van der Waals Epitaxy of MoS$_2$ Layers Using Graphene As Growth Templates. *Nano Lett.* **2012,** *12*, 2784-2791.

(19) Hong, S.; Kim, E. S.; Kim, W.; Jeon, S.-J.; Lim, S. C.; Kim, K. H.; Lee, H.-J.; Hyun, S.; Kim, D.; Choi, J.-Y.; Lee, Y. H.; Baik, S., A Hybridized Graphene Carrier Highway for Enhanced Thermoelectric Power Generation. *Phys. Chem. Chem. Phys.* **2012,** *14*, 13527-13531.

(20) Shibata, T.; Takano, H.; Ebina, Y.; Kim, D. S.; Ozawa, T. C.; Akatsuka, K.; Ohnishi, T.; Takada, K.; Kogure, T.; Sasaki, T., Versatile van der Waals Epitaxy-Like Growth of Crystal Films using Two-Dimensional Nanosheets as a Seed Layer: Orientation Tuning of SrTiO$_3$ Films along Three Important Axes on Glass Substrates. *J. Mater. Chem. C* **2014,** *2*, 441-449.

(21) Woo, S.; Jeong, H.; Lee, S. A.; Seo, H.; Lacotte, M.; David, A.; Kim, H. Y.; Prellier, W.; Kim, Y.; Choi, W. S., Surface Properties of Atomically Flat Poly-Crystalline SrTiO$_3$. *Sci. Rep.* **2015,** *5*, 8822.

(22) Nakada, K.; Ishii, A., Migration of Adatom Adsorption on Graphene using DFT Calculation. *Solid State Commun.* **2011,** *151*, 13-16.

(23) Lee, S. A.; Jeong, H.; Woo, S.; Hwang, J.-Y.; Choi, S.-Y.; Kim, S.-D.; Choi, M.; Roh, S.; Yu, H.; Hwang, J.; Kim, S. W.; Choi, W. S., Phase Transitions via Selective Elemental Vacancy Engineering in Complex Oxide Thin Films. *Sci. Rep.* **2016,** *6*, 23649.

(24) Niu, F.; Wessels, B. W., Surface and Interfacial Structure of Epitaxial SrTiO$_3$ Thin Films on (0 0 1) Si Grown by Molecular Beam Epitaxy. *J. Cryst. Growth* **2007,** *300*, 509-518.

(25)     Choi, W. S.; Rouleau, C. M.; Seo, S. S. A.; Luo, Z.; Zhou, H.; Fister, T. T.; Eastman, J. A.; Fuoss, P. H.; Fong, D. D.; Tischler, J. Z.; Eres, G.; Chisholm, M. F.; Lee, H. N., Atomic Layer Engineering of Perovskite Oxides for Chemically Sharp Heterointerfaces. *Adv. Mater.* **2012,** *24*, 6423-6428.

(26)     Yu, Z.; Ramdani, J.; Curless, J. A.; Overgaard, C. D.; Finder, J. M.; Droopad, R.; Eisenbeiser, K. W.; Hallmark, J. A.; Ooms, W. J.; Kaushik, V. S., Epitaxial Oxide Thin Films on Si (001). *J. Vac. Sci. Technol., B* **2000,** *18*, 2139-2145.

(27)     Wei, Y.; Hu, X.; Liang, Y.; Jordan, D. C.; Craigo, B.; Droopad, R.; Yu, Z.; Demkov, A.; Edwards, J. L.; Ooms, W. J., Mechanism of Cleaning Si (100) Surface using Sr or SrO for the Growth of Crystalline $SrTiO_3$ Films. *J. Vac. Sci. Technol., B* **2002,** *20*, 1402-1405.

(28)     Liu, L.; Ryu, S.; Tomasik, M. R.; Stolyarova, E.; Jung, N.; Hybertsen, M. S.; Steigerwald, M. L.; Brus, L. E.; Flynn, G. W., Graphene Oxidation: Thickness-Dependent Etching and Strong Chemical Doping. *Nano Lett.* **2008,** *8*, 1965-1970.

(29)     Tenne, D. A.; Farrar, A. K.; Brooks, C. M.; Heeg, T.; Schubert, J.; Jang, H. W.; Bark, C. W.; Folkman, C. M.; Eom, C. B.; Schlom, D. G., Ferroelectricity in Nonstoichiometric $SrTiO_3$ Films Studied by Ultraviolet Raman Spectroscopy. *Appl. Phys. Lett.* **2010,** *97*, 142901.

(30)     Hilt Tisinger, L.; Liu, R.; Kulik, J.; Zhang, X.; Ramdani, J.; Demkov, A. A., Ultraviolet-Raman Studies of $SrTiO_3$ Ultrathin Films on Si. *J. Vac. Sci. Technol., B* **2003,** *21*, 53-56.

# Supporting Information

# Highly oriented SrTiO$_3$ thin film on graphene substrate

*Sang A Lee[†,‡,⊥], Jae-Yeol Hwang[§,⊥], Eun Sung Kim[§], Sung Wng Kim[*,∥], and Woo Seok Choi[*,†]*

[†]Department of Physics, Sungkyunkwan University, Suwon, 16419, Korea

[‡]Institute of Basic Science, Sungkyunkwan University, Suwon, 16419, Korea

[§]Center for Integrated Nanostructure Physics, Institute for Basic Science (IBS), Suwon 16419, Korea

[∥]Department of Energy Sciences, Sungkyunkwan University, Suwon 16419, Korea

\* E-mail: kimsungwng@skku.edu (S.W.K.)

\* E-mail: choiws@skku.edu (W.S.C.)

[⊥]These authors contributed equally to this work.

**EXPERIMENTAL PROCEDURES**

**1. Thin film growth and structural characterization.**

**1a. Graphene growth.**

Graphene was synthesized on a chemical-mechanical polished (CMP) Cu foil (70 $\mu$m thickness, Nilaco Co., Tokyo, Japan) in a 3-in quartz tube through the thermal chemical vapor deposition (CVD) method. First, the Cu foil was heated up to 1060°C for 40 min and annealed for 90 min under $H_2$ atmosphere. After the annealing process, $H_2$ and $CH_4$ gases were injected into the furnace to synthesize graphene at 1060°C for 5 min. Finally, the Cu foil was rapidly cooled down to room temperature without any gas flow. Then, the few-layer-graphene film on the Cu foil was transferred onto an *a*-$SiO_2$ (amorphous phase with 300 nm thickness)/Si substrate by successive wet etching processes, including coating of poly(methyl methacrylate) (PMMA; MicroChem, e-beam resist, 950 k C4), etching (Transene, Danvers, MA, CE-100) of residual Cu, and removal of PMMA using acetone. The sample size of graphene on the *a*-$SiO_2$/Si substrate was ~10×10 $mm^2$.

**1b. $SrTiO_3$ growth.**

$SrTiO_3$ thin films were grown on *a*-$SiO_2$/Si single-crystalline substrates with and without a graphene by using pulsed laser deposition (PLD) at 700°C. A laser (248 nm; Lightmachinery, IPEX 864) fluence of 2 $J/cm^2$ and repetition rate of 3 Hz were used. The oxygen partial pressure was $10^{-6}$ Torr. Low oxygen partial pressure was employed in order to prevent oxidization of the graphene layer. We found that higher temperature or oxygen partial pressure resulted in the formation of polycrystalline phase in the $SrTiO_3$ thin films. Details of the chamber configuration can be found elsewhere.[26] The atomic structure and the crystal orientations of the $SrTiO_3$ thin films were characterized by high resolution x-ray diffraction (XRD). The thickness of the SrTiO$_3$ thin film was determined through x-ray reflectometry (XRR) as being ~40 ± 1 nm.

**2. Transmission electron microscopy (TEM).**

We prepared the sample foil by using the conventional method, which entailed mechanical thinning to ~10 $\mu$m and subsequent ion beam milling to electron transparency at an acceleration voltage of 0.5–3.5 kV using an Ar ion beam. The atomic strcuture was observed using a TEM instrument (JEOL JEM-ARM200F, JEOL, Ltd., Japan). The probe diameter of the beam was ~0.9 Å.

**3. Raman spectroscopy and resistivity measurements.**

The Raman spectra of all samples were acquired using a Renishaw Raman spectrometer (Renishaw inVia micro-Raman spectrometer) with an excitation wavelength of 532 nm. All the spectra are normalized to the Si peaks at ~960 cm$^{-1}$. Resistivity was measured as a function of temperature ($\rho(T)$) by using a low-temperature closed-cycle refrigerator. The measurements were performed from 300 K to 20 K by using the Van der Pauw method with In electrodes and Au wires.

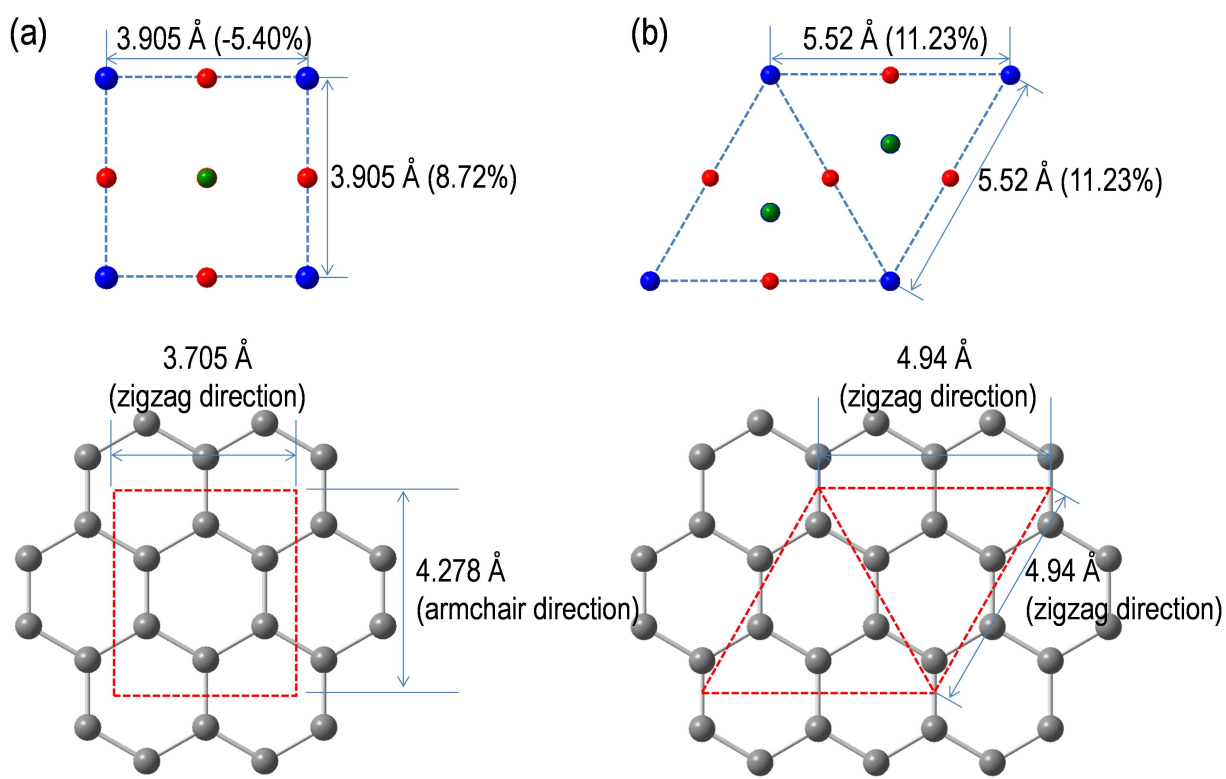

**Figure S1**. Simple lattice mismatch calculation of (a) (001)- and (b) (111)-oriented SrTiO$_3$. The epitaxial strain imposed by graphene on (001) SrTiO$_3$ thin film is ~8.72% (armchair direction) and −5.40% (zigzag direction) considering simple rectangular box on top of hexagonal network. The lattice mismatch of (111) SrTiO$_3$ thin film is ~11.23% (zigzag direction).

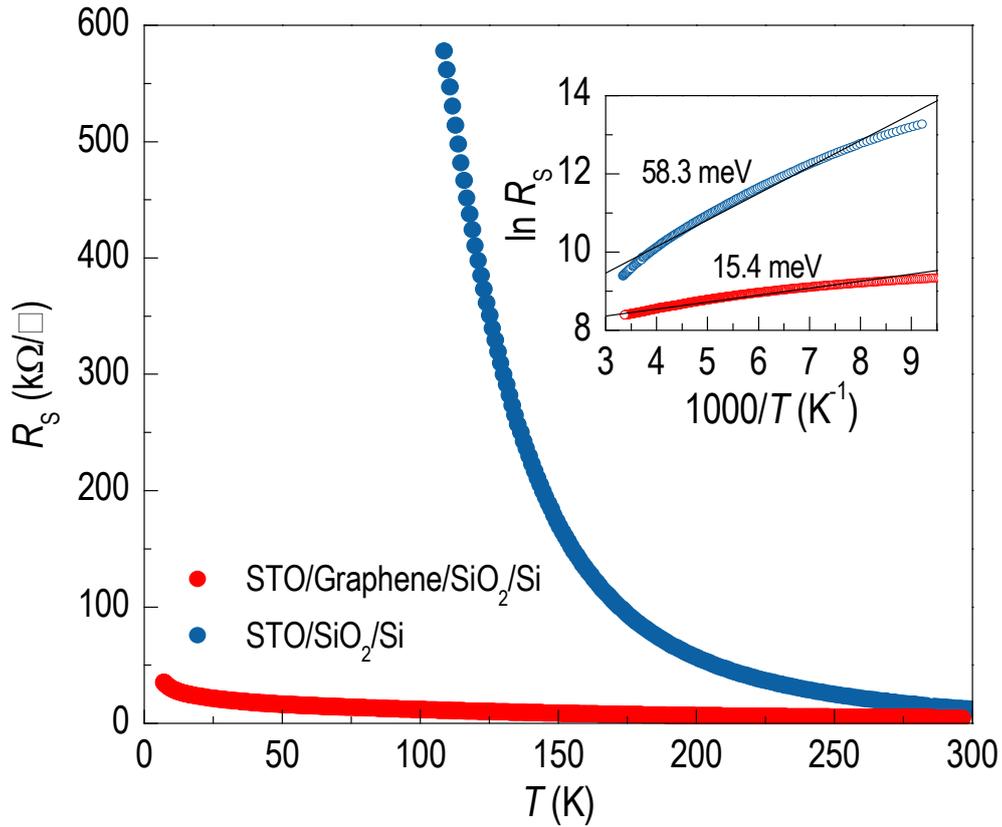

**Figure S2**. Temperature-dependent sheet resistance ($R_s(T)$) of SrTiO$_3$ thin films with and without graphene. The inset shows the Arrhenius plot of the electric conduction through SrTiO$_3$ with and without the graphene layer. The calculated activation energies are 58.3 and 15.4 meV for SrTiO$_3$ without and with graphene, respectively.

The graphene substrate induces lower resistance in the SrTiO$_3$ thin film by establishing the crystal structure as well as enabling electric conduction. Figure S2 shows the sheet resistance of the thin films as a function of temperature. Since the SrTiO$_3$ thin films were grown at low oxygen partial pressure, abundant oxygen vacancies could be induced, which resulted in the generation of free charge carriers. However, the resistivities reveal semiconducting temperature dependence, which might be due to the disorder induced by the grain boundaries. Nevertheless, the SrTiO$_3$ thin film on graphene substrate showed lower electric resistance over the whole temperature range studied compared to that without graphene. The activation energy obtained from Arrhenius plot was also lower (from 58.3 to 15.4 meV), as shown in the inset of Figure S2. These behaviors might originate from further enhanced carrier scattering by atomic disorders in the amorphous SrTiO$_3$ thin film. The presence of the intermediate layer in the SrTiO$_3$ thin film without graphene could also be detrimental for the electric conduction. On the other hand, the weak bonding nature facilitating the partial epitaxy between SrTiO$_3$ thin film and graphene substrate seems to be beneficial for the delocalized 3$d^1$ electrons in the Ti band of SrTiO$_3$ to flow.